\documentclass[journal,twoside,web]{ieeecolor}
\usepackage{tmi}
\usepackage{cite}
\usepackage{amsmath,amssymb,amsfonts}
\usepackage{algorithmic}
\usepackage{graphicx}
\usepackage{textcomp}
\usepackage{booktabs}
\usepackage{multirow}
\usepackage{subcaption}
\usepackage{hyperref}

\def\BibTeX{{\rm B\kern-.05em{\sc i\kern-.025em b}\kern-.08em
T\kern-.1667em\lower.7ex\hbox{E}\kern-.125emX}}
\begin{document}
\title{Filter2Noise: A Framework for Interpretable and Zero-Shot Low-Dose CT Image Denoising}
\author{Yipeng Sun, Linda-Sophie Schneider, Siyuan Mei, Jinhua Wang, Ge Hu, Mingxuan Gu, Chengze Ye, Fabian Wagner, Lan Song, Siming Bayer, and Andreas Maier, Member, IEEE
\thanks{This work was supported by the German Federal Ministry of Education and Research (BMBF) under the grant "Verbundprojekt 05D2022 - KI4D4E: Ein KI-basiertes Framework für die Visualisierung und Auswertung der massiven Datenmengen der 4D-Tomographie für Endanwender von Beamlines. Teilprojekt 5." Grant number: 05D23WE1. (Corresponding author: Yipeng Sun.)}
\thanks{Y. Sun, L.-S. Schneider, S. Mei, M. Gu, C. Ye, S. Bayer, and A. Maier are with the Pattern Recognition Lab, Friedrich-Alexander-University Erlangen-Nuremberg, 91058 Erlangen, Germany (e-mail: yipeng.sun@fau.de; linda-sophie.schneider@fau.de; siyuan.mei@fau.de; mingxuan.gu@fau.de; chengze.ye@fau.de; siming.bayer@fau.de; andreas.maier@fau.de).}
\thanks{F. Wagner is with Siemens Healthineers AG, 91301 Forchheim, Germany (e-mail: wagner.fabian@siemens-healthineers.com).}
\thanks{J. Wang and L. Song is with Department of Radiology, State Key Laboratory of Complex Severe and Rare Diseases, Peking Union Medical College Hospital, Chinese Academy of Medical Sciences and Peking Union Medical College, Beijing, China (e-mail: B2023001259@pumc.edu.cn; songl@pumch.cn).}
\thanks{G. Hu is with Theranostics and Translational Research Center, National Infrastructures for Translational Medicine, Institute of Clinical Medicine, Peking Union Medical College Hospital, Chinese Academy of Medical Sciences and Peking Union Medical College, Beijing, China (e-mail: huge@pumch.cn).}
}
\maketitle

\begin{abstract}
Noise in low-dose computed tomography (LDCT) can obscure important diagnostic details. While deep learning offers powerful denoising, supervised methods require impractical paired data, and self-supervised alternatives often use opaque, parameter-heavy networks that limit clinical trust. We propose Filter2Noise (F2N), a novel self-supervised framework for interpretable, zero-shot denoising from a single LDCT image. Instead of a black-box network, its core is an Attention-Guided Bilateral Filter, a transparent, content-aware mathematical operator. A lightweight attention module predicts spatially varying filter parameters, making the process transparent and allowing interactive radiologist control. To learn from a single image with correlated noise, we introduce a multi-scale self-supervised loss coupled with Euclidean Local Shuffle (ELS) to disrupt noise patterns while preserving anatomical integrity. On the Mayo Clinic LDCT Challenge, F2N achieves state-of-the-art results, outperforming competing zero-shot methods by up to 3.68 dB in PSNR. It accomplishes this with only 3.6k parameters—orders of magnitude fewer than competing models, which accelerates inference and simplifies deployment. By combining high performance with transparency, user control, and high parameter efficiency, F2N offers a trustworthy solution for LDCT enhancement. We further demonstrate its applicability by validating it on clinical photon-counting CT data. Code is available at \url{https://github.com/sypsyp97/Filter2Noise}.
\end{abstract}

\begin{IEEEkeywords}
Computed tomography (CT), deep learning, denoising, interpretable AI, self-supervised learning.
\end{IEEEkeywords}

\section{Introduction}
\label{sec:introduction}
\IEEEPARstart{L}{ow-dose} Computed Tomography (LDCT) is widely used in medical imaging to provide three-dimensional anatomical information for various diagnostic applications. Its implementation is guided by the "As Low As Reasonably Achievable" (ALARA) principle, which aims to minimize patient exposure to ionizing radiation \cite{chen2017low, Kolditz2010Low}. However, reducing the radiation dose increases quantum and electronic noise in the acquired projection data, which propagates and becomes structured during tomographic reconstruction. From an information theory perspective \cite{shannon1948mathematical}, this noise reduces the signal-to-noise ratio (SNR), making it more difficult to distinguish the true anatomical signal. This reduction in SNR can obscure subtle pathological features, such as low-contrast hepatic lesions, hairline fractures, or early-stage tumors, thereby lowering image quality and potentially compromising diagnostic accuracy \cite{wagner2023noise2contrast}. Consequently, the development of effective and reliable image denoising algorithms is important for improving the clinical utility and safety of LDCT.

For decades, traditional algorithms have served as the foundation of image denoising. Established methods such as Non-Local Means (NLM) \cite{buades2011nonlocal} and Block-Matching and 3D Filtering (BM3D) \cite{dabov2007image} are grounded in patch similarity, with BM3D further utilizing collaborative filtering within the transform domain. While effective under certain conditions, they often struggle in clinical settings. Their high computational demands can be prohibitive, and their performance is highly sensitive to hand-tuned parameters that must be adjusted for different noise levels and image content. Furthermore, they rely on explicit noise models (e.g., additive white Gaussian noise) that fail to accurately represent the complex, spatially correlated noise found in LDCT images \cite{wang2022uformer,hendriksen2020noise2inverse}.

The advent of deep learning has revolutionized this field \cite{tian2020deep, izadi2023image, archana2024deep}. Supervised methods, often employing powerful architectures like U-Net or Transformers \cite{wang2022uformer,ronneberger2015u}, learn complex, non-linear mappings from noisy to clean images. When trained on sufficient data, these methods often achieve state-of-the-art performance. However, their primary and most significant limitation is the requirement for large-scale datasets of perfectly registered noisy and clean image pairs. Acquiring such data in a clinical LDCT context is ethically challenging and practically infeasible, as it would necessitate exposing patients to both a low radiation dose and a full diagnostic dose simultaneously for the sole purpose of data collection \cite{Zhou2020Supervised,lu2023machine}.

To overcome this data dependency, self-supervised learning has emerged as a powerful alternative. The Noise2Noise (N2N) \cite{lehtinen2018noise2noise} framework demonstrated that a network can learn to denoise using only pairs of noisy images of the same scene, provided the noise is zero-mean. Subsequent methods have focused on the more practical single-image scenario, using techniques like pseudo-pair generation \cite{moran2020noisier2noise, hendriksen2020noise2inverse}, subsampling (Neighbor2Neighbor, NB2NB) \cite{huang2021neighbor2neighbor}, or "blind-spot" networks (Noise2Void, N2V) \cite{krull2019noise2void,batson2019noise2self,wang2022blind2unblind}. While innovative, these methods face two major hurdles to clinical adoption. First, many still require large datasets of noisy images to train robust models, making them vulnerable to domain shift when applied to data from different scanners or protocols. This data scarcity is amplified for emerging modalities like photon-counting CT (PCCT) \cite{alves2024deep}, where large datasets do not yet exist, creating a need for data-efficient, zero-shot denoising solutions. Second, they almost universally employ deep, complex architectures (e.g., U-Net) that function as black boxes \cite{castelvecchi2016can,batson2019noise2self,wang2022blind2unblind,lee2022ap,wang2023noise2info}. This lack of transparency is a significant barrier to clinical adoption. A clinician cannot easily understand, verify, or adjust the model's behavior, eroding the trust that is essential for high-stakes decisions related to patient safety and diagnostic confidence \cite{petch2022opening}. A non-transparent algorithm that may subtly alter or remove a critical diagnostic feature is a risk many practitioners are unwilling to take \cite{huang2018some}.

This motivates the search for a hybrid approach that combines the performance of data-driven methods with the interpretability and reliability of traditional operators \cite{wurfl2016deep}. Among these traditional methods, the bilateral filter  \cite{tomasi1998bilateral} stands out for its edge-preserving properties and has been widely explored for CT denoising \cite{zeng2022performance, wagner2022ultralow}. However, its classic implementation is limited by the need for globally fixed, manually selected filter parameters, which are not optimal for the varied content within each CT slice. A single slice may contain uniform soft tissue, high-contrast bone, and air-filled cavities, all requiring distinct filtering strategies. While some works have made these filter parameters learnable \cite{wagner2022ultralow}, the resulting filtering is still applied uniformly across the entire image, failing to adapt to local anatomical context.

In this paper, we address these fundamental challenges by proposing Filter2Noise (F2N), a framework designed to address the trade-off between performance and clinical trust in LDCT denoising. Our approach differs from conventional deep networks. Instead of a black box with millions of unconstrained parameters, the core of F2N is our proposed Attention-Guided Bilateral Filter (AGBF), a fully differentiable and inherently interpretable denoising operator. This philosophy aligns with the principle of learning with known operators, which are known to improve sample efficiency and robustness by constraining the solution space \cite{maier2019learning,wurfl2016deep}. We take a significant step forward from our prior work: rather than learning globally fixed parameters, F2N employs a lightweight attention module that learns to predict \textit{spatially varying} filter parameters tailored to the local content of the input image. The crucial distinction from a black-box U-Net is that the network's output is not an image, but a small set of physically meaningful, visualizable parameters that directly control a known, well-understood operator. This makes the overall process verifiable and controllable. Since this filter-based operator cannot learn abstract data representations in the same way a deep network can, making it trainable from a single noisy image requires a dedicated approach. We therefore introduce a novel self-supervised strategy that combines a multi-scale loss function with a unique data augmentation technique, the Euclidean Local Shuffle (ELS), which is specifically designed to break the spatial correlations in LDCT noise that often cause other self-supervised models to fail. The main contributions of this work are summarized as follows:

\begin{enumerate}
    \item We introduce \textbf{Filter2Noise (F2N)}, a new paradigm for zero-shot CT denoising that is interpretable by design. At its core is the Attention-Guided Bilateral Filter (AGBF), which replaces the black-box network with a transparent, mathematically defined operator whose local parameters are predicted by a lightweight attention module.
    \item We propose a novel self-supervised zero-shot training strategy while handling spatially correlated noise in a single image. This strategy combines a multi-scale loss with our Euclidean Local Shuffle (ELS) technique, which disrupts noise patterns while preserving anatomical integrity.
    \item We establish a new state-of-the-art for zero-shot denoising on the Mayo Clinic LDCT challenge, outperforming competing methods by up to 3.68 dB in PSNR. This is achieved with only 3.6k parameters, orders of magnitude fewer than deep learning models.
    \item We empower clinicians with direct control over the denoising process. The framework allows for the visualization and interactive, post-training adjustment of the learned filter parameters, enabling region-specific refinement to enhance diagnostic confidence.
    \item We validate the clinical readiness and generalization of F2N on real-world photon-counting CT data, demonstrating its ability to elevate low-dose images to a quality statistically indistinguishable from full-dose scans, proving its utility for next-generation imaging.
\end{enumerate}

\section{Methodology}
\label{sec:method}

Self-supervised zero-shot denoising methods often operate by constructing a pair of noisy training images from a single noisy input, typically via downsampling strategies \cite{mansour2023zero,huang2021neighbor2neighbor,lequyer2022fast}. This allows the application of a learning objective inspired by the Noise2Noise principle \cite{lehtinen2018noise2noise}, which can be generally formulated as:
\begin{equation}
\min_{\theta} \mathbb{E}_{y \sim p_{\text{noisy}}} \left[ \| f_{\theta}(g_1(y)) - g_2(y) \|^2 \right],
\label{eq:n2n}
\end{equation}
where \(y\) is the single noisy image, \(g_1\) and \(g_2\) are distinct downsampling operators that generate two different noisy views of the same underlying scene, and \(f_{\theta}\) is a neural network with parameters $\theta$. However, this paradigm faces two critical challenges in the context of clinical LDCT. First, the network-based approach, where $f_{\theta}$ is typically a deep U-Net, functions as a black box \cite{mansour2023zero,huang2021neighbor2neighbor,lequyer2022fast}, offering no insight into its decision-making process and thus hindering clinical trust. Second, the highly correlated noise inherent in reconstructed CT images can cause the network to learn a trivial identity mapping instead of effective denoising, as the input and target signals remain too similar.

To address these issues, the Filter2Noise (F2N) framework replaces the black-box network with our Attention-Guided Bilateral Filtering (AGBF) as its core denoising mechanism. This operator-based approach is paired with a self-supervised training strategy resilient to correlated noise. The following sections detail the AGBF operator and the F2N training strategy.

\subsection{Attention-Guided Bilateral Filter (AGBF)}

The AGBF is a differentiable and adaptive modification of the classic bilateral filter, designed for interpretable denoising. Standard bilateral filters \cite{tomasi1998bilateral} use fixed, global parameters, which must be manually tuned and are ill-suited to the heterogeneous anatomical content and spatially varying noise within a single CT slice. Wagner et al. \cite{wagner2022ultralow} improved upon this by making the parameters of the classical bilateral filter trainable, removing the need for manual tuning, but the learned parameters were still applied globally, failing to adapt to local image content. AGBF addresses this by rendering the filter parameters spatially dependent and conditioned on the input image content. This allows the filter to dynamically adjust its smoothing behavior based on local image features and noise levels, applying strong filtering in uniform regions while preserving sharp edges elsewhere.

\begin{figure*}[t]
	\centering
	\includegraphics[width=0.95\linewidth]{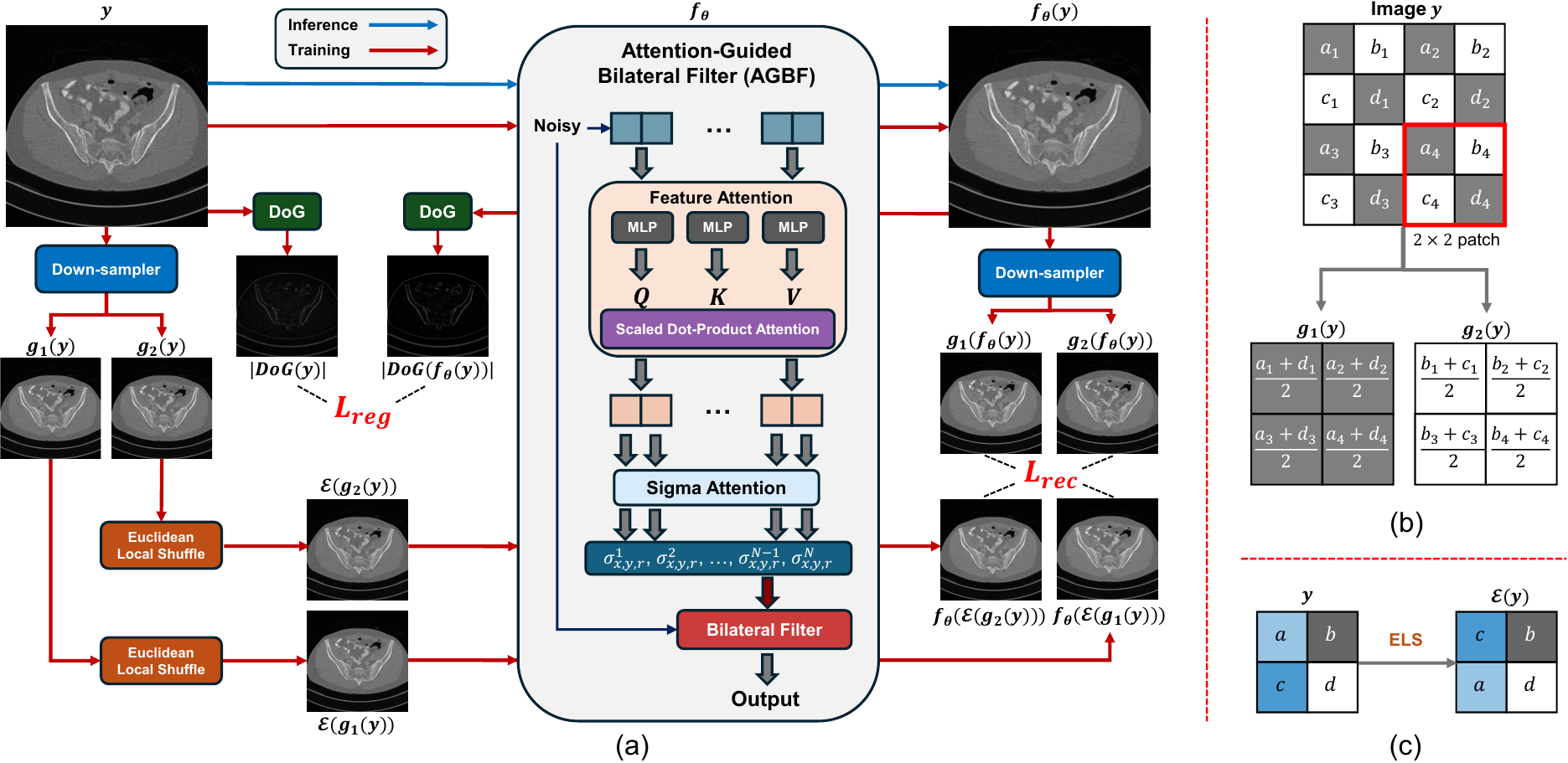}
	\caption{An overview of the Filter2Noise (F2N) framework. (a) The complete denoising pipeline. A single noisy input image \(y\) is used to generate two downsampled noisy views, \(g_1(y)\) and \(g_2(y)\). Our proposed Euclidean Local Shuffle (ELS) is applied to both views to decorrelate noise patterns before being processed by the multi-stage Attention-Guided Bilateral Filter (AGBF) network. The model is trained self-supervisedly with a multi-scale loss function that ensures consistency across different scales and transformations. (b) The downsampling process, based on ZS-N2N \cite{mansour2023zero}, which uses two distinct, non-overlapping convolutional kernels to create two distinct noisy sub-images from the single input. (c) The Euclidean Local Shuffle (ELS) technique. For each \(2 \times 2\) block of pixels, ELS identifies the pair of pixels with the smallest intensity difference and swaps their positions. This disrupts spatially correlated noise while preserving the local statistics of the underlying image structure.}
	\label{fig:method}
\end{figure*}

Figure \ref{fig:method}a depicts the architecture of AGBF. At its core is a lightweight dual-attention module that estimates the spatially varying parameters $\sigma_{r}^{(n)}$, $\sigma_{x}^{(n)}$, and $\sigma_{y}^{(n)}$ (range and spatial standard deviations) for each image patch at every denoising stage $n$. In contrast to conventional bilateral filters that apply uniform smoothing across the entire image, AGBF dynamically adjusts to local image characteristics, thereby achieving greater spatial adaptivity.

To provide context for our modifications, we first present an overview of the standard bilateral filter \cite{tomasi1998bilateral}. It computes a denoised pixel value \(f_\theta(y)_{x,y}\) as a weighted average of neighboring pixels, where \(y_{i,j}\) represents the intensity at the position $(i,j)$:
\begin{equation}
f_\theta(y)_{x,y} = \frac{1}{W_{x,y}} \sum_{i=x-k}^{x+k} \sum_{j=y-k}^{y+k} y_{i,j} \cdot w_s(i,j; x,y) \cdot w_r(i,j; y_{x,y}),
\label{eq:bf}
\end{equation}
where \(w_s\) and \(w_r\) are the spatial and range weighting kernels, respectively. These weights are determined by globally fixed spatial standard deviations (\(\sigma_x, \sigma_y\)) and a range standard deviation (\(\sigma_r\)). \(W_{x,y}\) is a normalization factor, and $k$ is the kernel size:
\begin{equation}
W_{x,y} =  \sum_{i=x-k}^{x+k} \sum_{j=y-k}^{y+k} w_s(i,j; x,y) \cdot w_r(i,j; y_{x,y}).
\end{equation}

The proposed AGBF (Figure \ref{fig:method}a) extends this formulation by replacing the fixed parameters with spatially varying values: $\sigma_{r}^{(n)}$, $\sigma_{x}^{(n)}$, and $\sigma_{y}^{(n)}$. These parameters are estimated for each small image patch through a lightweight dual-attention module. The core intuition behind this dual-attention design is that decoupling the tasks of content analysis and parameter prediction creates a more stable and effective learning dynamic. We separate these responsibilities into two specialized pathways: Feature Attention and Sigma Attention. The Feature Attention path is dedicated to understanding the semantic content of the image---distinguishing between different tissue types, such as soft tissue, bone, and air. The Sigma Attention path then takes this semantic understanding and focuses on the distinct task of mapping these features to the optimal, spatially varying filter parameters required for effective, content-aware denoising. This separation of concerns allows each module to specialize, leading to a more robust and interpretable model, a choice supported by our ablation study (see Table \ref{tab:ablation}), which shows that a single attention path performs worse.

In our experiments, a patch size of $P=8$ was empirically chosen to balance filter granularity and computational efficiency. Smaller patches (e.g., $P=4$) allow for finer detail but increase computational load and the model complexity required to build sufficient context, whereas larger patches (e.g., $P=16$) are computationally faster but result in less precise filtering, losing the ability to adapt to fine-grained local details. The input image \(x \in \mathbb{R}^{B \times C \times H \times W}\) is divided into non-overlapping \(P \times P\) patches, which are then flattened and processed by the dual-attention module.

\paragraph{Feature Attention:} This step is responsible for extracting rich, patch-wise features to understand the broader context of each local image region. It uses linear layers to project the flattened patch data into query (\(Q\)), key (\(K\)), and value (\(V\)) vectors:  
\begin{equation}
Q = W_q x_p, \quad K = W_k x_p, \quad V = W_v x_p,
\end{equation}
where \(x_p \in \mathbb{R}^{B \times N \times (P^2C)}\) is the flattened patch data, \(N = (H/P) \times (W/P)\) is the total number of patches, and \(W_q, W_k, W_v\) are learnable weight matrices. These vectors are then fed into a standard scaled dot-product attention mechanism \cite{vaswani2017attention}. The attention function is defined as:
\begin{equation}
\text{Attention}(Q, K, V) = \text{softmax}\left(\frac{QK^T}{\sqrt{d_k}}\right)V,
\label{eq:attention}
\end{equation}
where $d_k$ is the dimension of the key vectors. This mechanism computes a contextualized representation for each patch by aggregating information from all other patches, weighted by their similarity.

\paragraph{Sigma Attention:} This step predicts the filter's standard deviations (\(\sigma_{r}^{(n)}\), \(\sigma_{x}^{(n)}\), \(\sigma_{y}^{(n)}\)) for each patch, using the contextualized features generated by the Feature Attention pathway. A distinct set of linear layers (\(W_{q\sigma}, W_{k\sigma}, W_{v\sigma}\)) forms a new set of query (\(Q_\sigma\)), key (\(K_\sigma\)), and value (\(V_\sigma\)) vectors tailored specifically for this prediction task. These are processed through another attention layer (as defined in Eq. \ref{eq:attention}), followed by Layer Normalization and a final projection. For instance, the range standard deviation for a patch is calculated as follows:  
\begin{equation}
\sigma_r^{(n)} = \text{Softplus}\Bigl(W_{\sigma} \, \text{Attention}(Q_\sigma, K_\sigma, V_\sigma)\Bigr),
\end{equation}
where \(W_{\sigma}\) is a learnable matrix that projects the attention output to a scalar value, and the Softplus activation function ensures a non-negative constraint on the predicted standard deviation. The spatial standard deviations (\(\sigma_{x}^{(n)}, \sigma_{y}^{(n)}\)) are calculated following the same method with their own projection weights.

The predicted standard deviation values are then used to compute the spatially varying weighting kernels for each patch. At stage \(n\), the spatial weighting kernel is:
\begin{equation}
w_s^{(n)}(i,j; x,y) = \exp\left(-\frac{(i-x)^2}{2\left(\sigma_{x}^{(n)}\right)^2} - \frac{(j-y)^2}{2\left(\sigma_{y}^{(n)}\right)^2}\right),
\label{eq:spatial_kernel}
\end{equation}
and the range weighting kernel is:
\begin{equation}
w_r^{(n)}(i,j; x,y) = \exp\left(-\frac{\|y_{i,j} - y_{x,y}\|^2}{2\left(\sigma_{r}^{(n)}\right)^2}\right).
\label{eq:range_kernel}
\end{equation}
These adaptive kernels are then applied within Equation \ref{eq:bf} to obtain the denoised output \(f^{(n)}(y)_{x,y}\). The filter's kernel size \(k\) is also adapted dynamically for each patch at stage \(n\) according to the rule:
\begin{equation}
k = 2 \times \lceil \max(\sigma_{x}^{(n)}, \sigma_{y}^{(n)}) + 1 \rceil.
\end{equation}
This strategy rounds up the maximum predicted spatial standard deviation plus one, and doubles it for symmetry, ensuring the filter's receptive field is sufficiently large to accommodate the learned smoothing behavior while balancing computational cost and contributing to the model's efficiency.

\subsection{Interpretability and User Control}
A key advantage of AGBF is its interpretability. The denoising process is explicitly governed by the learned, spatially varying standard deviations (\(\sigma_{r}^{(n)}\), \(\sigma_{x}^{(n)}\), \(\sigma_{y}^{(n)}\)), allowing direct visualization of these parameter maps to gain insight into the model’s denoising behavior. For instance, a \(\sigma_r\) map would display high values in homogeneous areas such as the liver—indicating strong smoothing—and low values along sharp boundaries, like those of bones—signaling preservation of detail. In addition, AGBF empowers users to interactively adjust the standard deviation maps after training. This enables region-specific refinement; for example, a user can manually increase \(\sigma_{r}^{(n)}\) in areas suspected of containing low-contrast lesions, thereby enhancing both diagnostic confidence and interpretability. Furthermore, users can set upper bounds for the filter parameters to avoid excessive blurring in critical regions. These limits can be customized for particular anatomical areas, thus accommodating diverse noise characteristics or radiologist preferences tailored to specific diagnostic tasks.

\subsection{Training Strategy for Filter2Noise (F2N)} 

Unlike deep learning models that can leverage statistical information to learn abstract representations, our AGBF is a filter. This characteristic necessitates a dedicated training strategy and loss function to guide the learning process. The F2N framework (Figure \ref{fig:method}) leverages a multi-stage AGBF architecture. Our training strategy is designed to exploit image self-similarity across different scales and to robustly handle the spatially correlated noise common in LDCT. To achieve this, we introduce a multi-scale reconstruction loss, an edge preservation regularization term, and our proposed Euclidean Local Shuffle (ELS) technique.

\subsubsection{Downsampling} Following the approach of Zero-Shot Noise2Noise (ZS-N2N) \cite{mansour2023zero}, two downsampled images, \(g_1(y)\) and \(g_2(y)\), are generated from the noisy input \(y\). This is done using a \(2 \times 2\) convolution with a stride of 2, employing the following fixed kernels:
\begin{equation}
F_1 = \begin{bmatrix} 0 & 0.5 \\ 0.5 & 0 \end{bmatrix}, \quad
F_2 = \begin{bmatrix} 0.5 & 0 \\ 0 & 0.5 \end{bmatrix}.
\end{equation}
These kernels average different, non-overlapping pixel combinations from each $2 \times 2$ block of the original image, as illustrated in Figure \ref{fig:method}b, creating two distinct noisy observations of the same underlying anatomy.

\subsubsection{Euclidean Local Shuffle (ELS)}
To address the challenge of spatially correlated noise in low-dose CT, we present ELS (Figure \ref{fig:method}c), a technique grounded in Ruderman's principle of local statistical invariance in natural images \cite{ruderman1993statistics}. As we will demonstrate quantitatively in Table \ref{tab:els_validation}, this operation effectively decorrelates local noise patterns with minimal disruption to the underlying anatomical structure, a key property for robust self-supervised training.
While simple downsampling often struggles to fully decorrelate the noise, ELS actively disrupts local noise patterns by rearranging pixels within non-overlapping \(2 \times 2\) blocks. For each block, it swaps the pair of pixels with the minimum Euclidean distance in intensity, thereby maintaining local statistics while breaking the fine-grained noise correlation. For a \(2 \times 2\) block with pixel values (a, b, c, d), the pairwise Euclidean distances are calculated:
\begin{equation}
\begin{aligned}
d_{ab} &= \|a - b\|, \quad d_{ac} = \|a - c\|, \quad d_{ad} = \|a - d\|, \\
d_{bc} &= \|b - c\|, \quad d_{bd} = \|b - d\|, \quad d_{cd} = \|c - d\|.
\end{aligned}
\end{equation}
After determining the minimum distance \(d_{min} = \min(d_{ab}, d_{ac}, d_{ad}, d_{bc}, d_{bd}, d_{cd})\), the first corresponding pair of pixels is swapped. This operation, denoted as \(\mathcal{E}(y)\), is applied to both downsampled images, generating \(\mathcal{E}(g_1(y))\) and \(\mathcal{E}(g_2(y))\) as the final inputs for the training process.

\begin{table}[!t]
\centering
\caption{Validation of ELS Effectiveness in Noise Decorrelation. The metrics show ELS preserves content while reducing noise correlation, assessed by decomposing images into content (low-frequency) and noise (high-frequency) components.}
\label{tab:els_validation}
\begin{tabular}{lccc}
\toprule
\textbf{Metric} & \textbf{Baseline} & \textbf{After ELS} \\
\midrule
\multicolumn{3}{l}{\textit{\textbf{Content Correlation (Target: Preserve)}}} \\
Content Correlation & 1.0000 & 1.0000 \\
\midrule
\multicolumn{3}{l}{\textit{\textbf{Noise Correlation (Target: Destroy)}}} \\
Noise Correlation & 0.9835 & 0.9514 \\
Noise Frequency Correlation & 0.4750 & 0.3771 \\
Noise Local Correlation Mean & 0.8826 & 0.8431 \\
\bottomrule
\end{tabular}

\end{table}

\subsubsection{Loss Function} For self-supervised single-image training, the total loss \(L_{total}\) is a composite function that combines a multi-scale reconstruction loss, \(L_{rec}\), and a weighted regularization term \(\lambda L_{reg}\), where $\lambda$ is a weighting factor controlling the balance between denoising and detail preservation:
\begin{equation}
L_{total} = L_{rec} + \lambda L_{reg}.
\label{eq:loss}
\end{equation}
\paragraph{Reconstruction Loss (\texorpdfstring{\(L_{rec}\)}{L\_rec}):} This term enforces consistency across multiple resolutions and transformations, as illustrated in the pipeline in Figure~\ref{fig:method}a. It is composed of four distinct $L_1$ loss terms:
\begin{equation}
\begin{aligned}
L_{rec} = (
    \| f_\theta(\mathcal{E}(g_1(y))) - f_\theta(\mathcal{E}(g_2(y))) \|_1 \\
    + \| f_\theta(\mathcal{E}(g_1(y))) - g_1(f_\theta(y)) \|_1 \\
    + \| f_\theta(\mathcal{E}(g_2(y))) - g_2(f_\theta(y)) \|_1 \\
    + \| g_1(f_\theta(y)) - g_2(f_\theta(y)) \|_1
) / 3.
\end{aligned}
\end{equation}
The key idea behind this multi-faceted loss is to ensure that the underlying clean image content remains consistent and invariant across the various scales and transformations introduced during the self-supervised training process. Each component of the loss has a specific role in guiding the model. The first term, $\| f_\theta(\mathcal{E}(g_1(y))) - f_\theta(\mathcal{E}(g_2(y))) \|_1$, is the core self-supervised objective, enforcing that the denoised versions of the two different downsampled views should be identical, as they originate from the same underlying clean signal. The second and third terms, $\| f_\theta(\mathcal{E}(g_1(y))) - g_1(f_\theta(y)) \|_1$ and $\| f_\theta(\mathcal{E}(g_2(y))) - g_2(f_\theta(y)) \|_1$, enforce cross-scale consistency; they ensure that denoising a downsampled image yields the same result as denoising the full-resolution image and then downsampling it. This prevents scale-dependent artifacts. Finally, the fourth term, $\| g_1(f_\theta(y)) - g_2(f_\theta(y)) \|_1$, regularizes the final output by requiring the two different downsampled versions of the full-resolution denoised image to be consistent with each other, further promoting structural integrity.

\begin{table*}[!t]
\centering
\caption{Quantitative Comparison on the Mayo Low-Dose CT Challenge Datasets. Methods are categorized by different training paradigms. F2N-S1 and F2N-S2 denote Filter2Noise with one and two AGBF stages, respectively. Inference time is per slice on a consumer-grade NVIDIA RTX 4070 Super GPU. Paired t-tests compare each method with F2N-S2. {\textsuperscript{\dag}} indicates no statistically significant difference ($p > 0.05$). Best results for self-supervised methods are in bold.}
\label{tab:results}
\resizebox{\textwidth}{!}{%
\begin{tabular}{lcccccccc}
\toprule
\multirow{2}{*}{\textbf{Method}} & \multicolumn{2}{c}{\textbf{Mayo-2016 B30 (ID)}} & \multicolumn{2}{c}{\textbf{Mayo-2016 D45 (ID)}} & \multicolumn{2}{c}{\textbf{Mayo-2020 (OOD)}} & \multirow{2}{*}{\textbf{GPU Time (s)}} & \multirow{2}{*}{\textbf{\# Params.}} \\
\cmidrule(lr){2-3} \cmidrule(lr){4-5} \cmidrule(lr){6-7}
& \textbf{PSNR (dB) $\uparrow$} & \textbf{SSIM $\uparrow$} & \textbf{PSNR (dB) $\uparrow$} & \textbf{SSIM $\uparrow$} & \textbf{PSNR (dB) $\uparrow$} & \textbf{SSIM $\uparrow$} & & \\
\midrule
\multicolumn{9}{l}{\textit{\textbf{Supervised Method}}} \\
Noise2Clean \cite{ronneberger2015u} & 40.52 & 0.9250 & 39.83 & 0.9010 & 35.80 & 0.8750 & - & 2.2M \\
\midrule
\multicolumn{9}{l}{\textit{\textbf{Noisy Dataset-Based}}} \\
Noise2Void \cite{krull2019noise2void}
& 36.51 & 0.8915 & 36.28 & 0.8773 & 33.67 & 0.8421 & - & 2.2M \\
NB2NB \cite{huang2021neighbor2neighbor}
& 37.43 & 0.8945 & 37.91\textsuperscript{\dag} & \textbf{0.8997}\textsuperscript{\dag} & 34.97 & 0.8562 & - & 1.3M \\
\midrule
\multicolumn{9}{l}{\textit{\textbf{Zero-Shot}}} \\
BM3D \cite{dabov2007image}
& 37.15 & 0.8941 & 35.48 & 0.8780 & 36.50 & 0.8757 & 3 & -- \\
DIP \cite{ulyanov2018deep}
& 37.94 & 0.9074 & 36.23 & 0.8497 & 36.69 & 0.8592 & 180 & 2.2M \\
ZS-N2N \cite{mansour2023zero}
& 36.13 & 0.8733 & 38.01\textsuperscript{\dag} & 0.8924\textsuperscript{\dag} & 37.21\textsuperscript{\dag} & 0.8863 & 22 & 22k \\
\textbf{F2N-S1 (Ours)}
& 39.34 & 0.9126 & 37.59 & 0.8937 & 37.33\textsuperscript{\dag} & 0.8966\textsuperscript{\dag} & 8 & \textbf{1.8k} \\
\textbf{F2N-S2 (Ours)}
& \textbf{39.81} & \textbf{0.9154} & \textbf{38.14} & \text{0.8947} & \textbf{37.59} & \textbf{0.8977} & 16 & 3.6k \\
\bottomrule
\end{tabular}
}
\end{table*}

\begin{figure*}[!t]
\centering
\begin{subfigure}[b]{0.325\textwidth}
\centering
\includegraphics[width=\textwidth]{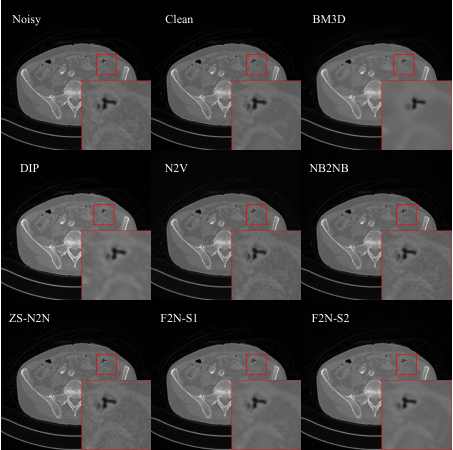}
\caption{Mayo-2016 B30}
\label{fig:b30}
\end{subfigure}
\hfill
\begin{subfigure}[b]{0.325\textwidth}
\centering
\includegraphics[width=\textwidth]{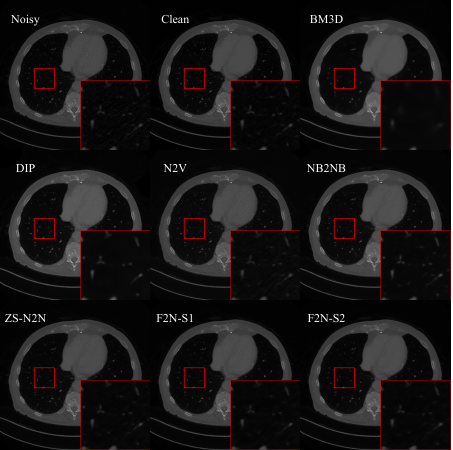}
\caption{Mayo-2016 D45}
\label{fig:d45}
\end{subfigure}
\hfill
\begin{subfigure}[b]{0.325\textwidth}
\centering
\includegraphics[width=\textwidth]{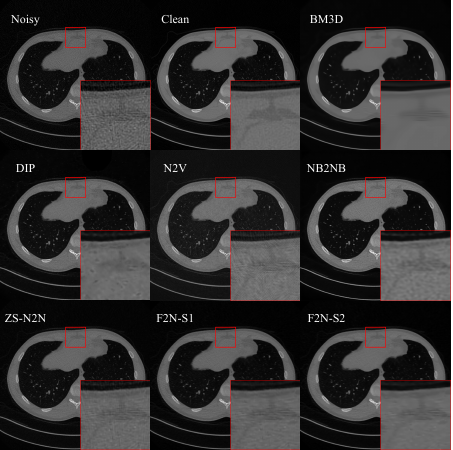}
\caption{Mayo-2020}
\label{fig:mayo2020}
\end{subfigure}
\caption{Qualitative comparison of denoising performance for self-supervised methods across different noise conditions. F2N effectively removes noise while preserving fine anatomical details. ZS-N2N struggles with the highly correlated noise in the B30 kernel image, leaving significant residual artifacts. BM3D tends to oversmooth fine details. F2N provides the best visual balance of noise reduction and structure preservation.}
\label{fig:results}
\end{figure*}

\paragraph{Regularization Loss (\texorpdfstring{\(L_{reg}\)}{L\_reg}):} This term is crucial for preserving sharp edges and preventing the excessive blurring that can occur with aggressive denoising. It penalizes differences between the absolute Difference of Gaussian (DoG) filtered versions of the noisy input and the denoised output:
\begin{equation}
    L_{reg} =  \left\| \, \left|\text{DoG}(y)\right| - \left|\text{DoG}(f_\theta(y))\right| \, \right\|_1,
\end{equation}
where the DoG operator is defined as the difference of two Gaussian convolutions:
\begin{equation}
    \text{DoG}(y) = (G_{s_2} * y) - (G_{s_1} * y).
\end{equation}
Here, $*$ denotes the 2D convolution operation, and $G_s$ is a 2D Gaussian kernel with standard deviation $s$, given by:
\begin{equation}
    G_s(u,v) = \frac{1}{2\pi s^2} \exp\left(-\frac{u^2+v^2}{2s^2}\right).
\end{equation}
We empirically set \(s_1 = 9\) and \(s_2 = 10\), as the model was not found to be highly sensitive to minor variations of these values. This specific configuration makes the DoG a close approximation of a Laplacian of Gaussian (LoG) filter, a robust edge detector, thereby enhancing the model's ability to preserve fine anatomical details while remaining robust to high-frequency noise.

\section{Experiments and Results}
\label{sec:experiments}

\subsection{Experimental Setup}

\subsubsection{Datasets}
To evaluate our zero-shot design, we conducted experiments on three distinct datasets: two public challenge datasets for in-domain (ID) and out-of-domain (OOD) evaluation, and one clinical dataset for real-world validation.

\paragraph{Mayo Clinic LDCT Datasets}
We used the 2016 and 2020 NIH AAPM-Mayo Clinic Low-Dose CT Grand Challenge datasets \cite{mccollough2017low,moen2021low}.
For \textbf{in-domain (ID) experiments}, the Mayo 2016 dataset was partitioned into six patients for training (3684 images), one for validation (L143, 585 images), and three for testing (L096, L109, L506; 1667 images). Images were reconstructed with both a smooth kernel (B30), which produces spatially correlated, textured noise, and a sharp kernel (D45), which results in finer, more random-like noise.
For \textbf{out-of-domain (OOD) experiments}, the Mayo 2020 dataset served as the test set. All dataset-based and supervised methods were trained exclusively on the Mayo 2016 B30 data and then evaluated on two patients from the Mayo 2020 set (C002, C004; 641 images), which were reconstructed using standard Filtered Backprojection (FBP) \cite{stierstorfer2004weighted}.
In total, our test set comprised 1667 slices for B30, 1667 for D45, and 641 for Mayo-2020. All images were $512 \times 512$ pixels with a 1 mm slice thickness.

\paragraph{Clinical Photon-Counting CT (PCCT) Data}
To evaluate our method's practical performance, we conducted a zero-shot validation on a clinical dataset from PUMCH. Data was acquired using a photon-counting CT scanner (Siemens Healthineers) and reconstructed with a QIR algorithm \cite{sartoretti2022quantum}. A patient was scanned at full and low doses, causing slight misalignments. We compared MTF-10\% and CNR in corresponding ROIs across 11 adjacent slices. This IRB-approved study had the patient's informed consent.

\subsubsection{Compared Methods}
We compared F2N against a suite of leading methods, categorized by their training paradigm:
\begin{itemize}
    \item \textbf{Zero-Shot Methods:} BM3D \cite{dabov2007image}, a high-performing classical algorithm; Deep Image Prior (DIP) \cite{ulyanov2018deep}, which uses network structure as a prior; and Zero-Shot Noise2Noise (ZS-N2N) \cite{mansour2023zero}. These are trained from scratch on each test image, like F2N.
    \item \textbf{Dataset-Based and Supervised Methods:} To establish performance benchmarks, we included Noise2Clean (N2C) \cite{ronneberger2015u}, a supervised U-Net approach serving as a practical upper bound, as well as Noise2Void (N2V) \cite{krull2019noise2void} and NB2NB \cite{huang2021neighbor2neighbor}.
\end{itemize}
Except for ZS-N2N, all dataset-based methods use the U-Net backbone \cite{ronneberger2015u}.

\subsubsection{Implementation Details}
We report Peak Signal-to-Noise Ratio (PSNR) and Structural Similarity Index (SSIM) as quantitative metrics, using full-dose images as ground truth. For the PCCT data, we report MTF-10\% spatial frequency and CNR. To assess statistical significance, comparisons against our final model (F2N-S2) were evaluated using paired t-tests (Table \ref{tab:results}). We tested F2N with one (F2N-S1) and two (F2N-S2) cascaded AGBF stages. The regularization weight $\lambda$ was selected from ten uniformly sampled values in $[200, 500]$ based on performance on the validation set. All models were trained for 500 epochs on one NVIDIA A100 80GB GPU using the AdamW optimizer with a learning rate of $1 \times 10^{-3}$.

\subsection{Results}

\subsubsection{Performance on Low-Dose CT Challenge Data}
Table \ref{tab:results} and Figure \ref{fig:results} summarize the quantitative and qualitative results. F2N consistently and significantly outperforms other zero-shot methods across all datasets and noise types.

On the \textbf{in-domain Mayo-2016 B30 dataset}, characterized by strong, correlated noise, F2N-S2 achieves a PSNR of 39.81 dB, a significant improvement of 1.87 dB over the next-best zero-shot method (DIP, 37.94 dB) and 3.68 dB over ZS-N2N (36.13 dB). This highlights the effectiveness of our ELS strategy in handling correlated noise, a common failure point for methods relying on simple subsampling. As Figure \ref{fig:b30} shows, ZS-N2N leaves significant residual noise, whereas F2N produces a much cleaner image while preserving fine structures.

On the \textbf{in-domain Mayo-2016 D45 dataset}, where noise is less correlated, the performance gap narrows, but F2N-S2 still achieves the highest PSNR and SSIM scores (38.14 dB). While NB2NB and ZS-N2N are competitive, F2N achieves this with a tiny fraction of the parameters: 3.6k for F2N-S2 versus 22k for ZS-N2N and 1.3M for NB2NB. This high parameter efficiency is a key advantage, enabling fast per-image optimization (16 seconds per slice on an NVIDIA RTX 4070 Super) and making deployment on standard clinical hardware feasible. The two-stage F2N-S2 consistently outperforms the single-stage F2N-S1, though F2N-S1, with only 1.8k parameters, already delivers strong performance in just 8 seconds per slice.

\begin{figure}[t]
\centering
\includegraphics[width=1.0\columnwidth]{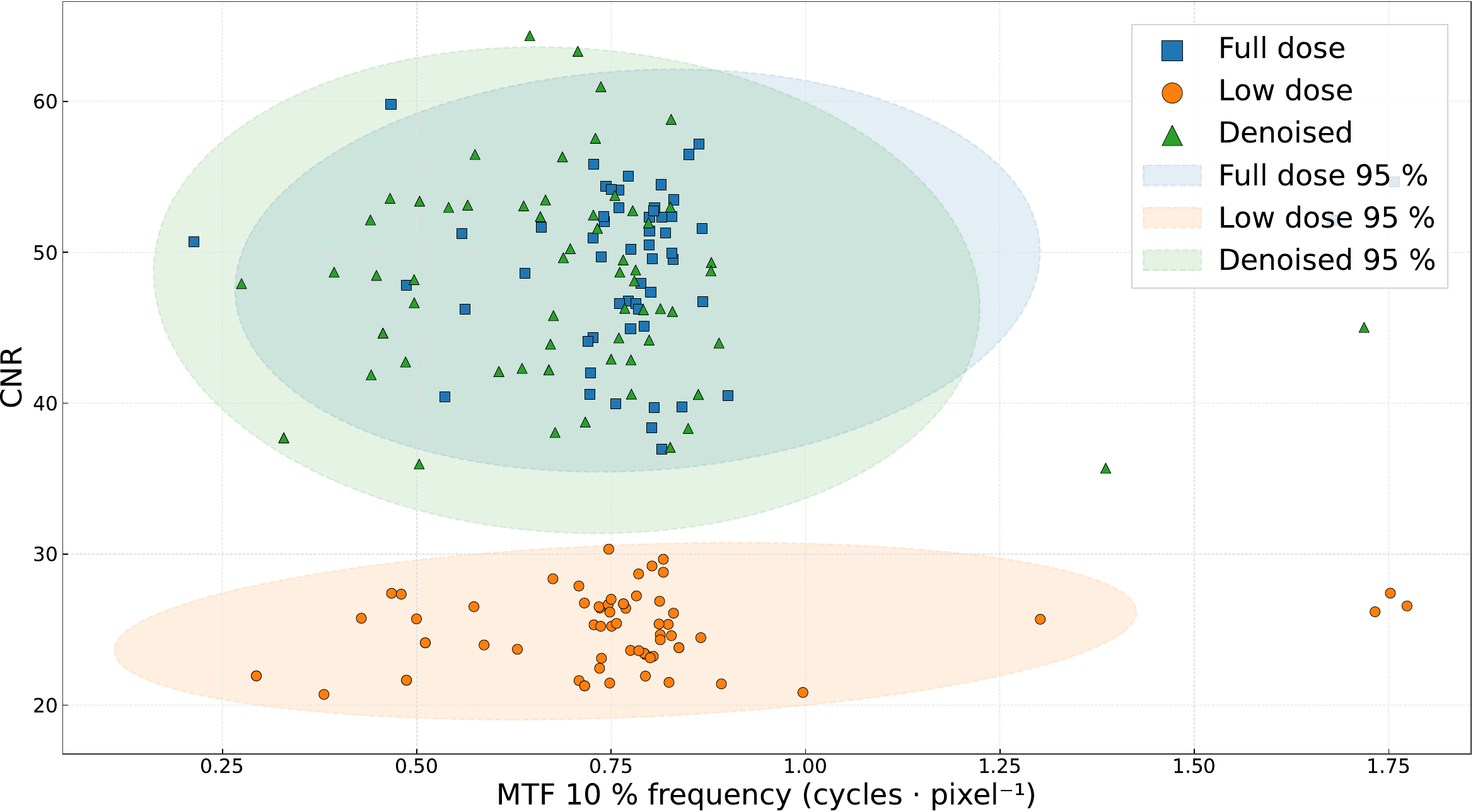} 
\caption{Zero-shot denoising performance on clinical PCCT data. Scatter plot of MTF-10\% spatial frequency versus contrast-to-noise ratio (CNR) for full-dose (blue squares), low-dose (orange circles), and denoised (green triangles) CT images. Each point represents a single measurement over corresponding regions of interest across 11 slices, while the shaded ellipses mark the bivariate 95\% confidence regions. The denoised data cluster closely with the full-dose reference in both resolution and CNR, whereas the low-dose data retain lower CNR and a slight reduction in MTF-10\% frequency.}
\label{fig:pumch_results}
\end{figure}

\subsubsection{Robustness to Domain Shift}
A critical advantage of the zero-shot paradigm is its inherent resilience to domain shift, a common failure point for methods reliant on large training datasets. This vulnerability is evident in our out-of-domain (OOD) evaluation on the Mayo-2020 dataset, as shown in Table~\ref{tab:results}. Methods trained on Mayo-2016 data, such as Noise2Clean and NB2NB, experience a significant performance decline when faced with this unfamiliar domain; Noise2Clean's PSNR plummets from 40.52 dB to 35.80 dB, and NB2NB's drops from 37.43 dB to 34.97 dB. In stark contrast, F2N, which optimizes on each image independently, is immune to this problem. It achieves a strong 37.59 dB on the Mayo-2020 data, underscoring its robust generalization and affirming its reliability for deployment across diverse clinical environments and scanner protocols.

\subsubsection{Validation on Clinical Photon-Counting CT Data}
To demonstrate F2N's utility on next-generation hardware, we analyzed its zero-shot performance on clinical PCCT data. The results, visualized in Figure~\ref{fig:pumch_results}, compare low-dose, denoised, and full-dose scans. The scatter plot of MTF-10\% spatial frequency versus CNR shows three distinct clusters. Low-dose scans (orange circles) exhibit a markedly lower CNR compared to the full-dose reference (blue squares), a statistically significant difference ($p < 10^{-88}$), while their MTF-10\% frequencies are not significantly different ($p = 0.50$).

F2N substantially improves image quality. The denoised data (green triangles) cluster closely with the full-dose reference, with their 95\% confidence ellipses largely overlapping. This indicates a successful restoration of image quality in terms of both noise and resolution. Paired t-tests confirm this visual assessment: F2N significantly boosted the CNR by a mean of 22.67 (95\% CI: [21.70, 23.64]), and critically, the CNR of the denoised images was not significantly different from the full-dose reference ($p = 0.10$). Furthermore, the MTF-10\% frequency was preserved, showing no statistical difference between the denoised and full-dose images ($p = 0.32$). In summary, F2N effectively elevates the diagnostic quality of low-dose PCCT images to a level statistically indistinguishable from full-dose scans.

\begin{figure}[!t]
\centering
\includegraphics[width=0.95\columnwidth]{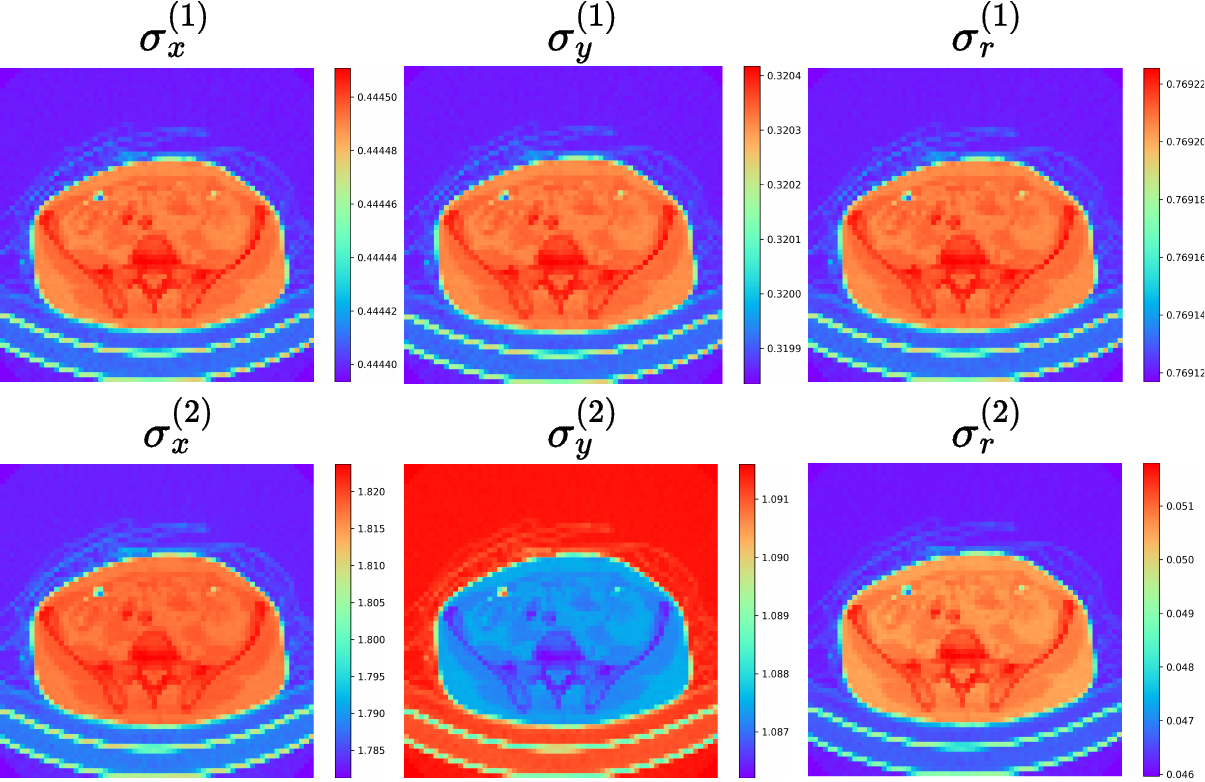}
\caption{Visualization of learned sigma maps from F2N-S2. The model learns to adapt the range (\texorpdfstring{$\sigma_r$}{sigma\_r}) and spatial (\texorpdfstring{$\sigma_x, \sigma_y$}{sigma\_x, sigma\_y}) filter parameters based on local image content. Maps for the same parameter type share a common color scale. This provides a clear interpretation of the adaptive denoising process.}
\label{fig:sigma_maps}
\end{figure}

\subsection{Interpretability and User Control in Practice}
A core contribution of F2N is its interpretability. Figure \ref{fig:sigma_maps} visualizes the spatially varying sigma maps predicted by F2N-S2 for an example image. The maps are shown without the underlying CT image for clarity, and maps of the same type (e.g., $\sigma_r$) share a common color scale for direct comparison between stages. The range sigma map ($\sigma_r$) reveals a counter-intuitive insight into the model's learned strategy. The model learns that high-contrast anatomical edges are inherently robust to smoothing. It can therefore apply aggressive denoising (high $\sigma_r$) to suppress the high-variance, correlated noise often found at these interfaces, relying on the large underlying intensity gradient to maintain edge sharpness. This demonstrates a nuanced understanding of the interplay between anatomy and noise characteristics, which is a hallmark of a content-aware system. This is a complex, non-obvious filtering strategy that a human-tuned or globally-fixed filter could not achieve, as it requires a local, content-aware understanding of the trade-off between noise characteristics and anatomical structure. This learned behavior provides evidence for the AGBF's capabilities. In contrast, lower $\sigma_r$ values are observed in regions with intermediate texture, where less aggressive filtering is required to preserve subtle details. The spatial sigma maps ($\sigma_x, \sigma_y$) demonstrate that the model learns anisotropic kernels, adapting the filter shape to local anatomical structures (e.g., elongating the kernel along boundaries). This visualization provides a direct, intuitive understanding of the denoising process, allowing clinicians to verify that the algorithm is behaving sensibly. The maps from Stage 2 show a general refinement of the Stage 1 maps, applying more subtle and localized adjustments.

\begin{figure}[!t]
\centering
\vspace{0.4em}
\includegraphics[width=0.90\columnwidth]{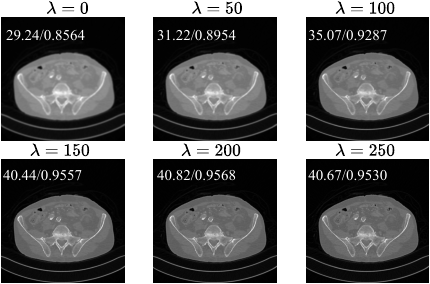}
\caption{Ablation study on the regularization weight \texorpdfstring{$\lambda$}{lambda} for the Mayo-2016 B30 dataset. The plot shows the trade-off between noise reduction and structural preservation. This highlights the importance of tuning \texorpdfstring{$\lambda$}{lambda} for the desired balance, a process that can be guided by visual inspection in a clinical workflow.}
\label{fig:ablation_study}
\end{figure}

\subsection{Ablation Studies}
\label{sec:ablation}
To validate our design choices, we conducted ablation studies on the challenging Mayo-2016 B30 dataset. The results, summarized in Table \ref{tab:ablation} and Figure \ref{fig:ablation_study}, systematically evaluate each component of F2N.

The component-wise analysis in Table \ref{tab:ablation} confirms that the loss function, architecture, and processing strategy are all critical for performance. Removing the multi-scale consistency objective from the loss results in a 2.36 dB PSNR drop. Architecturally, the attention-based module is crucial, outperforming a CNN alternative by 1.99 dB, and disabling spatial adaptivity leads to a 4.80 dB performance loss, underscoring its importance. The ELS technique is particularly vital for handling correlated noise, as its removal causes a 3.49 dB PSNR drop, a significant improvement over a random shuffle.

Finally, Figure \ref{fig:ablation_study} illustrates the impact of the regularization weight $\lambda$, highlighting a clear trade-off between noise reduction and detail preservation. 

\section{Discussion}
Our experiments show that Filter2Noise (F2N) addresses key limitations of existing denoising methods for LDCT. The framework's transparency is not an add-on, but a direct result of its design: F2N is interpretable because it \textbf{is a filter}, not a black-box approximator. The learned sigma maps are not a post-hoc \textit{explanation} of the model's behavior; they \textit{are} the behavior, offering a direct, verifiable view into the denoising mechanism. This stands in stark contrast to post-hoc eXplainable AI (XAI) methods, which can only provide an approximation of a deep network's reasoning and risk being unfaithful to the true decision-making process \cite{prakash2021interpretable,li2022interpretable}.

The post-training user control feature transforms F2N from a static tool into a dynamic, radiologist-guided diagnostic aid, enhancing clinical workflow and confidence. Radiologists can interactively adjust local smoothing—increasing it in uniform tissue to find low-contrast lesions or decreasing it over complex vasculature to preserve fine details. For example, when examining a scan for subtle liver metastases, a clinician can reduce the local smoothing strength ($\sigma$ parameter) in a specific region of interest. This targeted adjustment can "un-blur" a suspicious area, revealing finer textures and allowing for a more confident diagnosis.

\begin{table}[!t]
\centering
\caption{Component-wise ablation study on Mayo-2016 B30 dataset. Each row shows the impact of removing or modifying specific components of F2N-S2. The multi-scale loss and ELS are critical for handling correlated noise, while the attention mechanism provides superior parameter prediction compared to CNN alternatives.}
\label{tab:ablation}
\resizebox{\columnwidth}{!}{%
\begin{tabular}{lcc}
\toprule
\textbf{Configuration} & \textbf{PSNR (dB)} & \textbf{SSIM} \\
\midrule
\textbf{F2N-S2 (Full)} & \textbf{39.81} & \textbf{0.9154} \\
\midrule
\multicolumn{3}{l}{\textit{Loss Component Analysis}} \\
w/o main self-supervision term (term 1 of $L_{rec}$) & 37.45 & 0.8924 \\
w/o cross-scale consistency (terms 2 \& 3 of $L_{rec}$) & 38.21 & 0.9032 \\
w/o DoG edge preservation & 30.12 & 0.7998 \\
\midrule
\multicolumn{3}{l}{\textit{Architecture Variants}} \\
CNN Instead of Attention & 37.82 & 0.8967 \\
Single Attention (No Dual-Path) & 38.25 & 0.9004 \\
Fixed Global Parameters & 35.01 & 0.8743 \\
\midrule
\multicolumn{3}{l}{\textit{Processing Strategy}} \\
Three-Stage Processing & 39.68 & 0.9143 \\
w/o ELS (Simple Downsampling) & 36.32 & 0.8701 \\
Random Shuffle Instead of ELS & 37.88 & 0.8891 \\
\bottomrule
\end{tabular}%
}
\end{table}

Furthermore, our validation on clinical photon-counting CT data underscores the framework's versatility and future-readiness. PCCT is a promising technology that offers superior material differentiation but comes with its own unique noise characteristics and a scarcity of training data. The ability of F2N to significantly enhance image quality from a single low-dose PCCT scan, without prior training on such data, provides evidence of its robust generalization. This result suggests F2N is not only a solution for current LDCT challenges but also a tool for advancing emerging imaging modalities where data is inherently limited. A key advantage of the zero-shot paradigm employed by F2N is its inherent robustness to domain shift. While dataset-based methods can achieve good performance, they are vulnerable to performance degradation when applied to data from different scanners or protocols, as shown in Table \ref{tab:results}. F2N avoids this by optimizing on each image individually. This architectural choice also serves as a crucial \textbf{safety feature}. Because F2N is a constrained filter, it excels at restoring a signal but deliberately avoids inventing anatomical details where none exist. This "inability" to hallucinate is not a flaw but a core design principle that builds clinical trust by preventing the generation of plausible but false diagnostic information.

From a technical standpoint, the success of F2N, especially on the B30 kernel dataset with its correlated noise, validates our hypothesis that explicitly addressing this noise structure is critical for self-supervised single-image denoising. The proposed ELS technique proved to be a simple yet effective method for this purpose. The high parameter efficiency of F2N (1.8k-3.6k parameters) is another significant advantage. A model with only 3.6k parameters is far less likely to overfit to the noise in a single image compared to a model with millions of parameters, which further enhances its trustworthiness in a zero-shot setting. This makes it lightweight and fast to train on a per-image basis and opens the possibility of deployment on standard clinical workstations without dedicated GPUs, or even on CPUs for resource-constrained environments, lowering the barrier to clinical adoption.

\begin{figure}[!t]
\centering
\includegraphics[width=1.0\columnwidth]{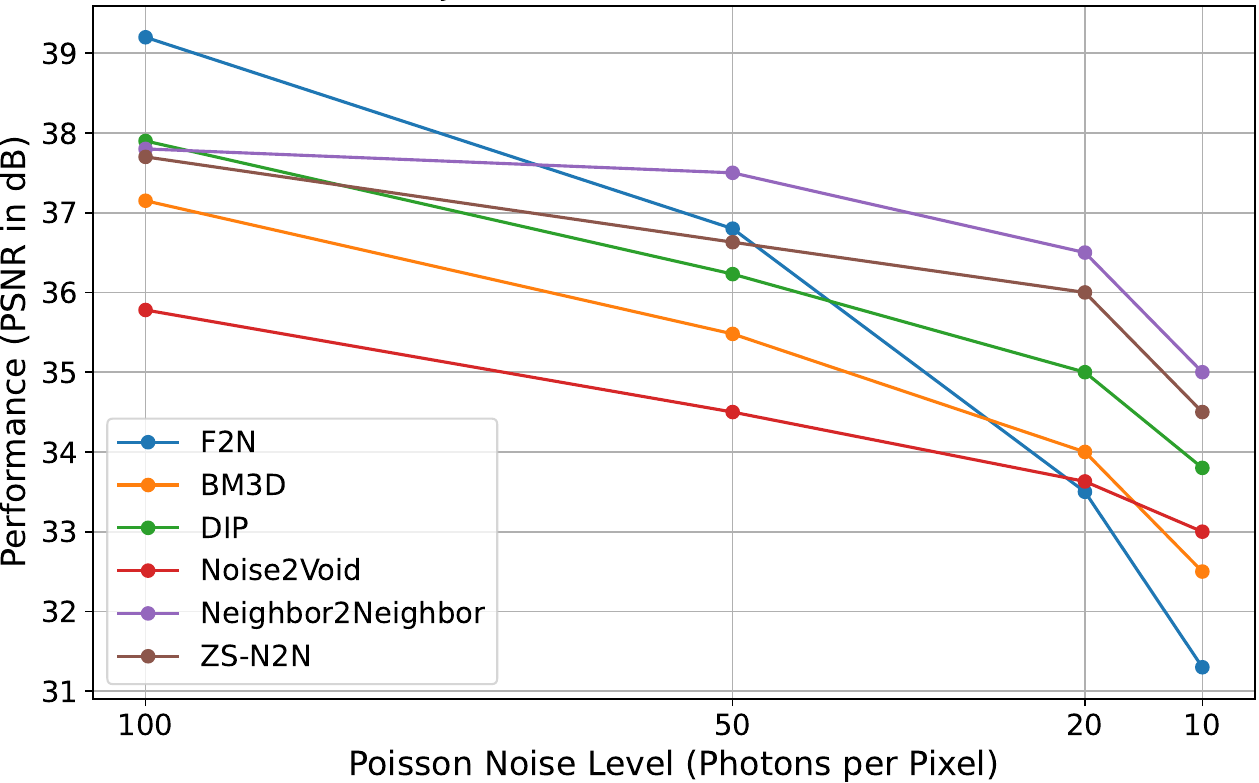}
\caption{Robustness analysis of different denoising methods across varying levels of Poisson noise. The performance, measured in PSNR (dB), is plotted against the number of photons per pixel, with noise levels increasing from left to right (from 100 down to 10 photons/pixel). The plot compares Filter2Noise (F2N) with BM3D, Deep Image Prior (DIP), Noise2Void, NB2NB, and ZS-N2N, showing the robustness of F2N across all tested noise intensities.}
\label{fig:robustness}
\end{figure}

\subsubsection*{Limitations and Future Work}

F2N operates as a filter rather than a generative model: it restores corrupted voxels but never hallucinates anatomy, lowering the chance of clinical artifacts at the cost of poorer recovery under extreme noise (Figure~\ref{fig:robustness}). The 16\,s per-slice test-time optimization is slower than a single forward pass, yet it is performed once per image and removes the substantial data and compute required to train a large network. This trade-off suits non-urgent tasks such as pre-operative planning, dose computation, and staging; faster variants would broaden use cases. Our ELS is still empirical, and its guarantees or failure modes are not yet characterized. Future work will automate the choice of~$\lambda$ and extend F2N to MRI, ultrasound, and other modalities.

\section{Conclusion}
In this work, we introduced Filter2Noise (F2N), a framework designed to address the gap between performance and clinical viability in LDCT denoising. It replaces opaque deep networks with our lightweight, interpretable Attention-Guided Bilateral Filter (AGBF) and enables zero-shot, self-supervised training via a multi-scale loss and our Euclidean Local Shuffle (ELS) technique to handle correlated noise. F2N significantly outperforms existing zero-shot methods in challenging scenarios while being orders of magnitude more parameter-efficient. Its combination of high performance, interpretability, user control, and robustness to domain shift makes F2N a practical solution for LDCT enhancement. Successful validation on clinical PCCT data further demonstrates its potential for next-generation imaging technologies.


\bibliographystyle{IEEEtran}
\bibliography{mybib}

\end{document}